\documentclass[a4paper,11pt]{article}

\usepackage{geometry}
\usepackage{graphicx}
\usepackage{amsmath,amssymb}
\usepackage[hidelinks]{hyperref}
\usepackage{booktabs}
\usepackage{siunitx}
\usepackage{microtype}
\usepackage{caption}
\usepackage{subcaption}
\usepackage{authblk} 
\usepackage{cite}
\usepackage{url}
\usepackage{orcidlink}

\title{\textbf{Pulse Shape Discrimination for Germanium Detectors using Variational Quantum Circuits}}

\author[1,2]{Fabrizio Napolitano\thanks{Electronic address: \texttt{fabrizio.napolitano@unipg.it}}\orcidlink{0000-0002-8686-5923}}
\affil[1]{Universit\`a di Perugia, Dipartimento di Fisica e Geologia, Via A. Pascoli, 06123 Perugia, Italy}
\affil[2]{INFN, Sezione di Perugia, Via A. Pascoli, 06123 Perugia, Italy}

\date{\today}

\begin{document}

\maketitle

\begin{abstract}
Pulse shape discrimination (PSD) is a critical component in background rejection for neutrinoless double-beta decay and dark matter searches using Broad Energy Germanium (BEGe) detectors. To date, advanced discrimination has relied on Deep Learning approaches employing e.g. Denoising Autoencoders (DAE) and Convolutional Neural Networks (CNN). While effective, these models require tens of thousands of parameters and heavy pre-processing. In this work, we present, to the best of our knowledge, the \textbf{first application of Quantum Machine Learning (QML) to real, experimental pulse waveforms} from a germanium detector. We propose a quantum-classical hybrid approach using Variational Quantum Circuits (VQC) with amplitude encoding. By mapping the 1024-sample waveforms directly into a 10-qubit Hilbert space, we demonstrate that a VQC with only \textbf{302 trainable parameters} achieves a receiver operating characteristic (ROC) area under the curve (AUC) of \textbf{0.98} and a global accuracy of \textbf{97.1\%}. This result demonstrates that even in the current Noisy Intermediate-Scale Quantum (NISQ) era, quantum models can match the performance of state-of-the-art classical baselines while reducing model complexity by over two orders of magnitude. Furthermore, we envision a scenario where future quantum sensors transmit quantum states directly to such processing units, exploiting the exponentially large Hilbert space in a natively quantum pipeline.
\end{abstract}

\section{Introduction}

Broad Energy Germanium (BEGe) detectors are a key technology in rare-event physics, widely employed in experiments searching for Beyond Standard Model physics, neutrinoless double-beta decay ($0\nu\beta\beta$) and Dark Matter, such as VIP, GERDA, and LEGEND \cite{Piscicchia2018,Agostini2011}. In these low-background environments, the sensitivity is limited by radioactive backgrounds that can mimic the signal of interest. Consequently, Pulse Shape Discrimination (PSD) is crucial to distinguish localized Single Site Events (SSE), characteristic of signal interactions, from Multi Site Events (MSE) typical of Compton-scattered background gamma rays. 

Historically, PSD relied on parametric approaches like the A/E method, which compares the maximum current amplitude to the total energy \cite{Budjas2009}. However, at low energies ($<10$ keV), the discrimination power of such parameters degrades significantly as electronic noise and microphonic disturbances begin to dominate the signal morphology. To overcome this, recent advances have pivoted toward Deep Learning (DL). Pipelines combining Denoising Autoencoders (DAE) for signal cleaning and Convolutional Neural Networks (CNN) for classification have established new state-of-the-art performance \cite{manti2025github, Holl2019}. Classical models achieve high fidelity, but they come at a computational cost: they often require tens of thousands of parameters and rely on heavy pre-processing steps to reconstruct pulse shapes before classification can occur. This complexity poses challenges for deployment in low-latency trigger systems or on-detector electronics where memory and power are constrained.

In this work, we introduce a compact quantum-classical pipeline based on Variational Quantum Circuits (VQC) \cite{Benedetti2019}. Quantum Machine Learning (QML) has recently gained traction in High Energy Physics, showing promise in image-based jet tagging~\cite{elhag2024quantum}, particle tracking, and event reconstruction~\cite{di2024quantum}. The primary motivation for applying QML to detector data is the exponential scaling of the quantum state space. By mapping inputs into a high-dimensional Hilbert space, quantum models can capture complex, non-linear correlations that classical architectures typically require significant depth and over-parameterization to resolve. Yet, despite this theoretical potential, the application of QML to \textit{raw} time-series data from nuclear and particle detectors remains largely unexplored. To the best of our knowledge, this is the first study benchmarking QML against classical DL using real, experimental waveforms from a BEGe detector.

Our approach uses \textit{amplitude encoding} to map 1024-sample time-bins directly into the amplitudes of a 10-qubit system, resulting in a logarithmic compression of the input feature space. Our key contributions are:
\begin{enumerate}
    \item[(i)] The development of an amplitude-encoded VQC that processes raw waveforms, eliminating the computational overhead of DAE pre-processing;
    \item[(ii)] The optimization of a strongly-entangling ansatz with only 302 trainable parameters, a reduction of two orders of magnitude compared to standard CNNs;
    \item[(iii)] A demonstration that NISQ-era quantum algorithms can match classical state-of-the-art performance in precision instrumentation tasks, paving the way for quantum-enabled detectors.
\end{enumerate}

\section{Methodology}

\subsection{Dataset and Pre-processing}
We make use of the public VIP-2 dataset \cite{ZenodoData}, consisting of 22376, 1024-sample current pulses from a BEGe detector, labeled by hand. 
The dataset is labeled into two classes:
\begin{itemize}
    \item Background (Label 0): Multi-Site Events (MSE) and noise artifacts.
    \item Signal (Label 1): Single-Site Events (SSE), which are the candidates for rare-event physics.
\end{itemize}
The class balance is approximately 85\% signal and 15\% background, reflecting the typical event rate after preliminary energy cuts.

\subsection{The Classical Benchmark Pipeline}
To establish a performance baseline, we refer to the recent work by Manti et al. \cite{manti2025github}, which establishes the current state-of-the-art for this specific dataset. Their approach addresses the low signal-to-noise ratio at low energies ($<10$ keV) through a two-stage Deep Learning pipeline:
\begin{enumerate}
    \item \textbf{Denoising Autoencoder (DAE):} A convolutional encoder-decoder network is first trained on a synthetic dataset of 5,000 pulses to suppress microphonic noise and electronic interference. The encoder compresses the waveform into a latent space of 128 dimensions using three convolutional layers (filters: 64, 32, 16).
    \item \textbf{CNN Classifier:} The reconstructed (cleaned) waveforms are then fed into a separate Convolutional Neural Network (CNN) for classification. This network comprises three sequential convolutional layers with increasing filter depth (32, 64, 128), followed by batch normalization, max-pooling, and a dense layer with dropout.
\end{enumerate}
The classical classifier typically utilizes a two-channel input (the waveform and its first derivative) to maximize feature extraction. This architecture achieves an Area Under the Curve (AUC) of 0.99 and an accuracy of 95\%. This pipeline is effective, but involves a significant number of trainable parameters ($>10^4$) and requires a dedicated pre-processing inference step (the DAE) before classification can occur.

\subsection{Quantum Pipeline: Pre-processing and Encoding}

Unlike the classical benchmark that employ trapezoidal filters and autoencoders for denoising, we apply a ``strict physics" minimal cleaning pipeline. This involves a baseline subtraction (calculated as the mean of the first 20 samples) followed by $L^2$ normalization. The normalization ensures the waveform vector $\mathbf{x}$ satisfies $\|\mathbf{x}\|_2 = 1$, a requisite for valid quantum state preparation. Importantly, \textbf{no denoising autoencoder is applied}, requiring the classifier to be robust against intrinsic electronic and microphonic noise.

\subsection{Amplitude Encoding}
To process high-dimensional time-series data on a quantum processor, efficient embedding is required. We employ amplitude encoding \cite{Mottonen2004, Schuld2020,weigold2020data}, which maps a real normalized vector $\mathbf{x} = (x_0, x_1, \dots, x_{N-1})$ into the amplitudes of a quantum state of $n = \log_2 N$ qubits.

Formally, the encoding prepares the state:
\begin{equation}
|\psi_x\rangle = \sum_{i=0}^{N-1} x_i |i\rangle,
\end{equation}
where $|i\rangle$ represents the computational basis state corresponding to the binary representation of index $i$.

\textbf{Example:} Consider a simplified waveform with 4 time-bins $\mathbf{x} = [0.5, 0.5, -0.5, 0.5]$. This requires $n=\log_2(4)=2$ qubits. The resulting quantum state is:
\[
|\psi\rangle = 0.5|00\rangle + 0.5|01\rangle - 0.5|10\rangle + 0.5|11\rangle.
\]
In our specific case, the input vector has dimension $N=1024$. This allows us to embed the entire time-series into just 10 qubits ($2^{10}=1024$). This logarithmic compression ($N \to \log N$) is a key advantage of the quantum framework, allowing small quantum registers to handle high-dimensional detector data. We note that the low qubit count, besides being compatible with current NISQ devices, also helps mitigate issues like barren plateaus during training \cite{McClean2018}.

While amplitude encoding offers an exponential compression of the input dimensionality, we acknowledge that its exact implementation on present-day hardware requires state preparation circuits of depth $\mathcal{O}(2^n)$ in the general case \cite{plesch2011quantum}. In the context of detector readout, however, several mitigating factors apply: (i) pulse waveforms exhibit strong temporal correlations and smoothness, allowing structured state-preparation approaches (such as tensor-network initializations or approximate loading) to achieve significantly lower depth than random state preparation \cite{araujo2021divide, holmes2020efficient}; (ii) the encoding circuits can be pre-compiled and optimized for the specific frequency response of the detector; and (iii) future quantum sensors may directly emit quantum states corresponding to the charge-collection dynamics \cite{khan2025quantum}, effectively bypassing the classical digitization and re-encoding bottleneck. These considerations suggest that amplitude encoding is not only theoretically motivated but also practically relevant for future quantum-native instrumentation pipelines.

\subsection{Variational Ansatz and Readout}
\begin{figure}[h!]
  \centering
  \begin{subfigure}[b]{0.90\textwidth}
    \includegraphics[width=0.95\linewidth,height=0.5\linewidth]{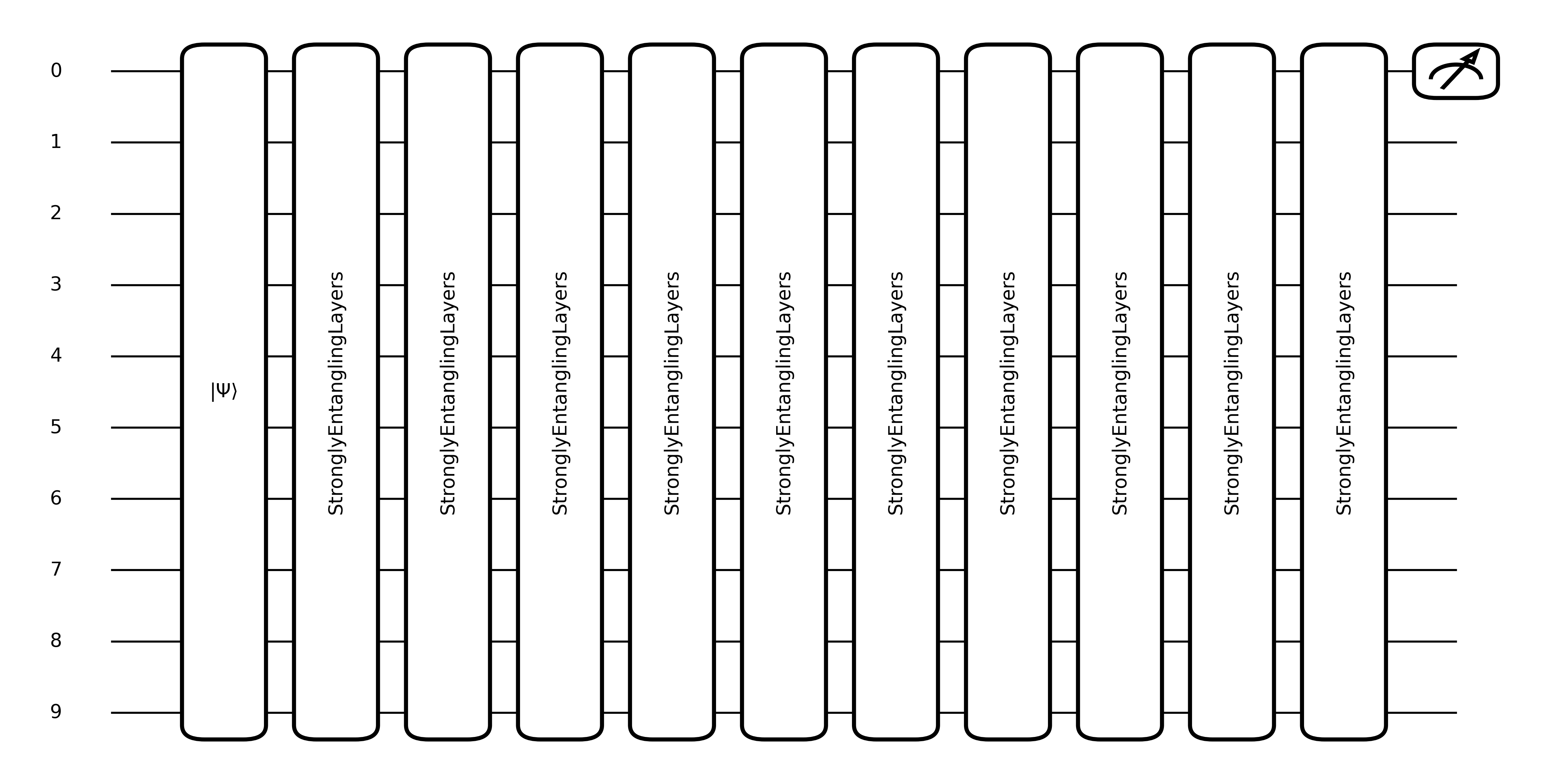}
    \caption{Full architecture of the Variational Quantum Circuit. The numbers on the left indicate qubit indices (0 to 9). The input waveform is amplitude-encoded into the 10-qubit register. The processing block consists of 10 layers of strongly entangling gates (rotations + CNOTs). The final measurement extracts the expectation value $\langle Z_0 \rangle$.}
  \end{subfigure}
  \hfill
  \begin{subfigure}[b]{0.90\textwidth}
    \includegraphics[width=0.95\linewidth,height=0.5\linewidth]{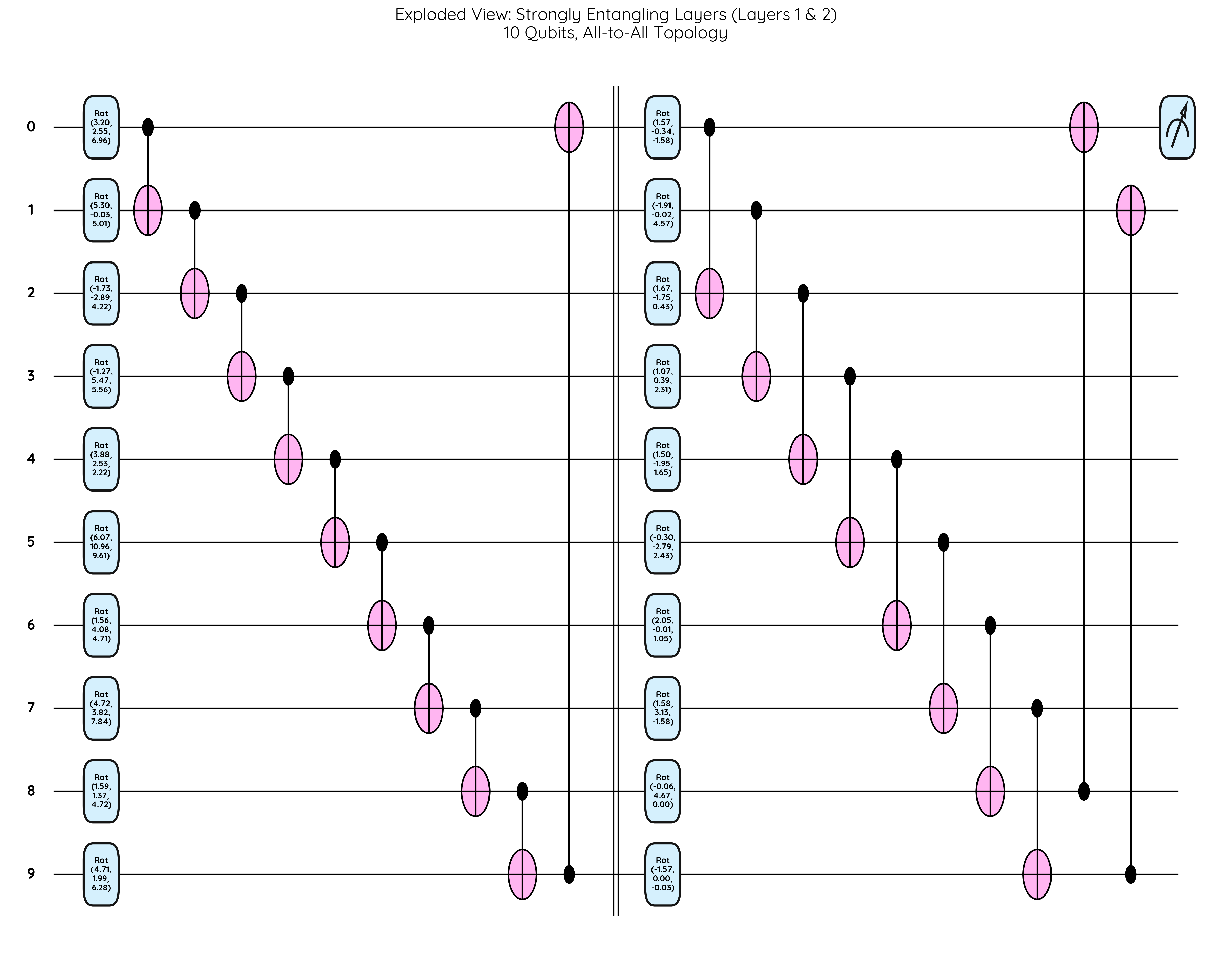}
    \caption{Detailed view of the last layer of the VQC. Each qubit undergoes a parameterized rotation followed by CNOT entangling gates, derived from~\cite{Schuld2020}}
  \end{subfigure}
  \caption{View of the Variational Quantum Circuit employed in this work. }
  \label{fig:architecture}
\end{figure}

The quantum model utilizes a layered ``Strongly Entangling" ansatz \cite{Schuld2020}, depicted in Figure \ref{fig:architecture}. Each layer $l$ consists of:
\begin{enumerate}
    \item \textbf{Rotations:} A generic single-qubit rotation $U3(\phi, \theta, \omega)$ applied to each qubit.
    \item \textbf{Entanglement:} A cascade of CNOT gates applied in a nearest-neighbor ring topology to distribute correlations across the register.
\end{enumerate}

For $L$ layers and $N_q$ qubits, the number of rotation parameters is $3 \times N_q \times L$. We select $N_q=10$ and optimize for $L=10$ layers. 

\textbf{The Readout Layer:}
The quantum circuit outputs the expectation value of the Pauli-$Z$ operator on the first qubit, $\langle Z_0 \rangle$. This value is bounded in $[-1, 1]$. To map this to a probability space suitable for binary classification, we apply a classical affine transformation followed by a sigmoid activation:
\begin{equation}
    y_{\text{pred}} = \sigma(\alpha \cdot \langle Z_0 \rangle + \beta)
\end{equation}
where $\alpha$ (scaling) and $\beta$ (bias) are trainable classical parameters. This final layer allows the model to learn the optimal decision boundary and sensitivity. The total parameter count is:
\begin{equation}
N_{\mathrm{params}} = (10 \times 10 \times 3) + 2 = 302.
\end{equation}

\subsection{Training Strategy}
Training deep parameterized quantum circuits is notoriously difficult due to the ``Barren Plateau" phenomenon, where the variance of the cost function gradients vanishes exponentially with the number of qubits and layers \cite{McClean2018}. A standard random initialization of a deep circuit ($L=10$) would likely result in an untrainable model where the optimizer is stuck on a flat landscape.

To circumvent this, we implement a \textit{Progressive Circuit Growing} strategy with identity-block initialization \cite{Skolik2021}. The training curriculum proceeds as follows:
\begin{enumerate}
	\item \textbf{Shallow Start:} We begin by optimizing a shallow circuit with depth $L=1$. This simplified landscape allows the optimizer to quickly find a coarse-grained solution.
	\item \textbf{Layer Expansion:} Upon convergence, we extend the circuit depth to $L+1$. The parameters for the first $L$ layers are initialized with the optimal values found in the previous stage (Warm Start).
	\item \textbf{Identity Initialization:} The parameters of the newly added layer are initialized close to zero ($\theta \sim \mathcal{U}(-10^{-3}, 10^{-3})$). Since the rotation gates are of the form $R(\theta)$, this initialization approximates the Identity operation. Consequently, the new deeper circuit starts with a functional behavior identical to the trained shallower circuit, preserving the learned feature map.
	\item \textbf{Joint Optimization:} All parameters (both old and new) are then unrolled and optimized jointly.
	\item \textbf{Repeat:} This process iterates until the target depth $L=10$ is reached.
\end{enumerate}
This approach ensures that the optimization trajectory always starts from a valid solution and gradually increases the expressibility of the ansatz, effectively guiding the solver through the Hilbert space without traversing the barren regions of the landscape.

\textbf{Epoch Schedule and Fine-Tuning:}
The training duration was adaptive to the circuit depth. For intermediate stages ($L=1 \dots 9$), the training was capped at 40 epochs; this ``structural assimilation" phase is sufficient to integrate the new layer without over-optimizing a sub-optimal architecture. 
Upon reaching the full target depth ($L=10$), we transitioned to a \textit{Global Convergence} phase consisting of 250 epochs. This extended duration allows the optimizer to fine-tune the correlations across the fully entangled 10-qubit register.

Furthermore, we applied a scheduled learning rate annealing. The base learning rate was decayed by a factor of 0.95 at each layer addition ($LR_L = 0.01 \times 0.95^{L-1}$) to ensure stability as the parameter space expanded. Within each stage, a ``Reduce-on-Plateau" scheduler monitored the validation accuracy, halving the learning rate if no improvement was observed for 30 epochs.

\subsection{Implementation and Computational Resources}
The quantum pipeline is implemented in Python using the \texttt{PennyLane} framework \cite{Bergholm2018}. To handle the simulation of the 10-qubit state vector efficiently, we used the \texttt{pennylane-lightning.gpu} plugin backed by the NVIDIA cuQuantum SDK. The optimization loop was built using \texttt{JAX} \cite{JAX2018} to enable Just-In-Time (JIT) compilation of the quantum gradients.

All simulations were performed on the \textbf{UniNuvola} cloud infrastructure provided by the University of Perugia/INFN. The specific computing instance was provisioned with:
\begin{itemize}
    \item \textbf{CPU:} 64 vCPUs derived from \textbf{AMD EPYC 9334} (32-Core) processors.
    \item \textbf{RAM:} 64 GB of system memory.
    \item \textbf{Accelerator:} A Multi-Instance GPU (MIG) slice of an \textbf{NVIDIA A30 Tensor Core GPU} (24 GB VRAM).
\end{itemize}
Despite the complexity of simulating a dense 10-qubit variational circuit with full entanglement, the JIT-compiled simulation completed the full progressive training curriculum (growing from 1 to 10 layers) in approximately \textbf{34 minutes}. This highlights that, unlike large classical models that often require hours of training on clusters, the proposed VQC is lightweight enough for rapid experimentation and deployment.

\section{Results}

\begin{figure}[t!]
  \centering
  \includegraphics[width=0.90\linewidth]{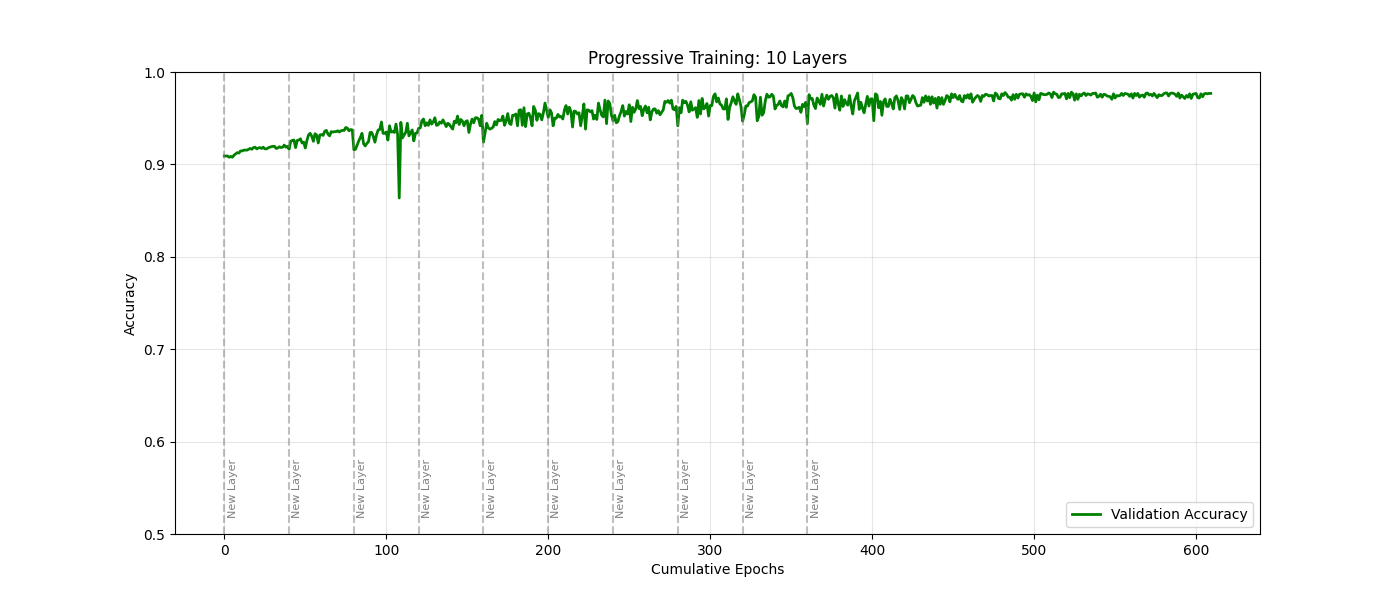}
  \caption{Validation accuracy across progressive training stages.}
  \label{fig:training}
\end{figure}

\begin{figure}[t!]
  \centering
  \includegraphics[width=0.90\linewidth]{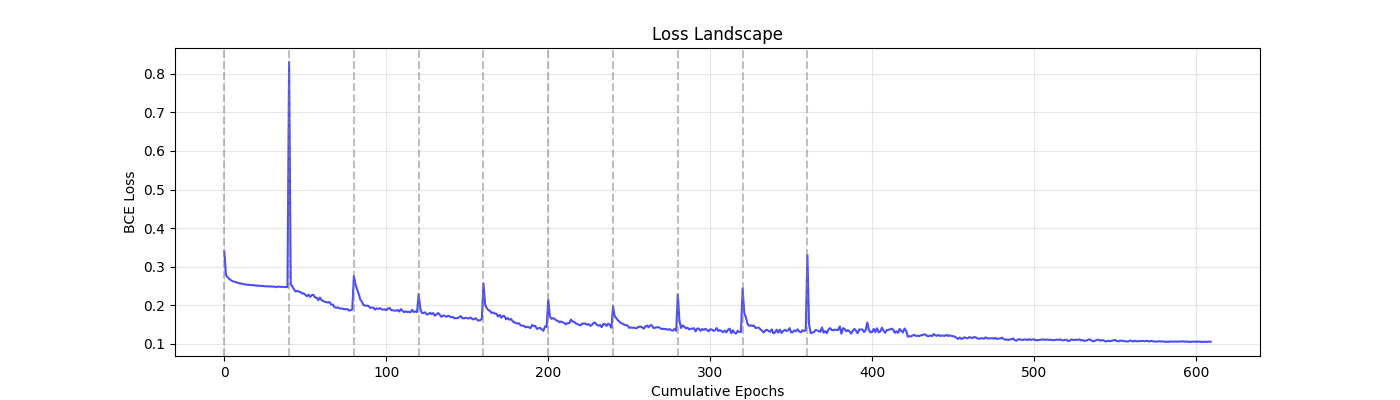}
  \caption{Loss across progressive training stages.}
  \label{fig:training_loss}
\end{figure}

\begin{figure}[t!]
  \centering
  \includegraphics[width=0.90\linewidth]{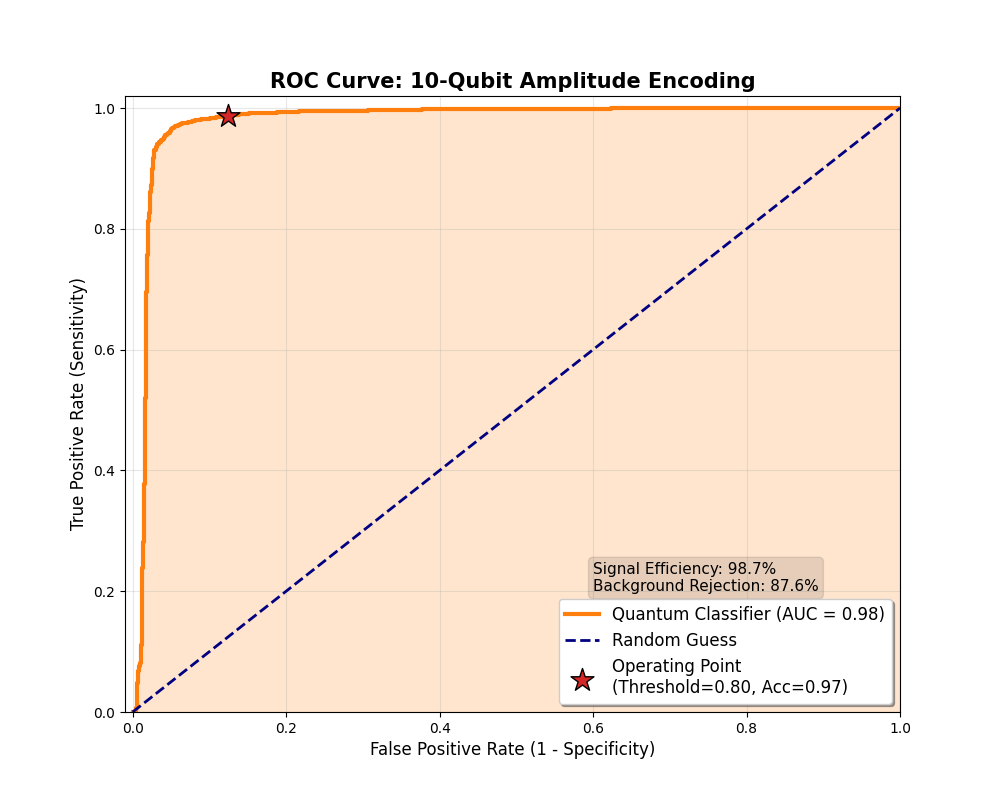}
  \caption{ROC curve illustrating the trade-off between true positive rate and false positive rate across different thresholds. The AUC of 0.98 indicates strong classification performance.}
  \label{fig:roc}
\end{figure}

\subsection{Classification Fidelity and Physics Performance}
The primary figure of merit for rare-event searches is the ability to maximize Signal Acceptance (retaining true single-site events) while maximizing Background Rejection (discarding multi-site events). The 10-qubit VQC was evaluated on the independent test set of 11,377 waveforms.

Figure \ref{fig:roc} displays the Receiver Operating Characteristic (ROC) curve. The model achieves an Area Under the Curve (AUC) of \textbf{0.98}, indicating exceptional separation power. The curve exhibits a steep initial rise, suggesting that the quantum model can reject a large fraction of background events with negligible signal loss.

We selected an operating threshold of $T=0.80$ by maximizing the F1-score on the validation set. At this working point, the model yields:
\begin{itemize}
    \item \textbf{Global Accuracy:} 97.1\%
    \item \textbf{Signal Efficiency (Recall):} \textbf{98.7\%} (9607/9732). This is a critical result for physics goals, as it ensures that less than 1.3\% of potential rare decays are accidentally vetoed by the algorithm.
    \item \textbf{Background Rejection:} \textbf{87.7\%} (1442/1645). The model successfully identifies the majority of multi-site Compton scatterings.
\end{itemize}
The confusion matrix confirms that the VQC has learned the morphological distinctions between the event classes directly from the raw amplitudes.

\subsection{Analysis of Progressive Training Dynamics}
The efficacy of the progressive layer-wise training strategy is visualized in Figure \ref{fig:training}. The plots reveal a distinct ``step-wise" improvement in model performance, correlating with the addition of quantum layers:
\begin{itemize}
    \item \textbf{Shallow Regime ($L=1-3$):} The model quickly reaches $\sim 94\%$ accuracy. This suggests that even a shallow quantum circuit with minimal entanglement can capture the gross features of the pulse (likely rise-time).
    \item \textbf{Deep Regime ($L=4-10$):} As depth increases, we observe finer improvements, pushing accuracy from 95\% to the final 97.1\%. This indicates that the deeper layers and increased entanglement capability allow the model to resolve subtle features, such as the small kinks in the rising edge typical of MSEs, which shallow circuits miss.
\end{itemize}

The validation accuracy (Figure \ref{fig:training}) shows no signs of divergence or catastrophic forgetting when new layers are added, validating the stability of the identity-block initialization and warm-start strategy.

\subsection{Benchmarking: The Efficiency Advantage}
Table \ref{tab:comparison} contrasts our Quantum pipeline with the classical State-of-the-Art (SotA) established by Manti et al. \cite{manti2025github}. 

While the classification metrics (AUC, Accuracy) are statistically equivalent, the parametric efficiency of the VQC is vastly superior. The classical benchmark relies on a Deep CNN (estimated $>50,000$ parameters) and a separate Denoising Autoencoder (DAE) for pre-processing. Our VQC achieves the same fidelity with only 302 parameters.
This represents a model compression factor of approximately $\mathbf{160\times}$. Furthermore, the VQC operates on raw waveforms, implying that the quantum feature map inherently performs the necessary noise filtering during the unitary transformation, removing the need for a dedicated DAE inference step.

\begin{table}[t]
	\centering
	\caption{Performance and resource comparison between the Classical SotA and the proposed Quantum VQC. The VQC matches the classical performance metric while using orders of magnitude fewer parameters and eliminating the pre-processing stage.}
	\begin{tabular}{lcc}
		\toprule
		\textbf{Metric} & \textbf{Classical SotA} \cite{manti2025github} & \textbf{This Work (VQC)} \\
		\midrule
		\textbf{Input Representation} & Waveform + Derivative & \textbf{Raw Waveform Only} \\
		\textbf{Noise Handling} & Denoising Autoencoder (DAE) & \textbf{Intrinsic (Quantum)} \\
		\textbf{Model Architecture} & Deep CNN (3 Layers + Dense) & \textbf{10-Qubit VQC (10 Layers)} \\
		\textbf{Trainable Parameters} & $\sim 50,000$ & \textbf{302} \\
		\midrule
		\textbf{ROC AUC} & 0.99 & \textbf{0.98} \\
		\textbf{Signal Efficiency} & 98\% & \textbf{98.7\%} \\
		\bottomrule
	\end{tabular}
	\label{tab:comparison}
\end{table}

\subsection{Stability and Generalization}
To ensure the result is not an artifact of a specific data split, the final evaluation was performed on a ``remainder" dataset of $>11,000$ pulses that were strictly excluded from the training and validation loops. The consistency between validation metrics (Acc 97.8\%) and test metrics (Acc 97.1\%) confirms that the 302-parameter model has learned generalizable physical features (such as charge collection time and current topology) rather than memorizing noise patterns, a common risk in over-parameterized classical networks. No overfitting was observed during training.

\section{Discussion}

\subsection{Extreme Compactness and Model Efficiency}
The most striking result of this study is the parametric efficiency of the VQC. The final model uses 10 layers (302 parameters), but the training dynamics (Figure \ref{fig:training}) reveal that the model becomes competitive much earlier. Already at Layer 6 ($\approx 182$ parameters), the validation accuracy exceeds 96.5\%. 
This observation has significant implications for deployment: (i) even as ``Quantum-Inspired" models on classical hardware, such compact architectures can be executed with minimal computational overhead; (ii) in strictly resource-constrained environments, such as radiation-hardened FPGAs or ASICs operating deep underground, a shallow 6-layer quantum model could provide ``good enough" discrimination with a memory footprint that is negligible compared to classical Deep Learning models and (iii) the low parameter count suggests that future implementations on NISQ hardware could be feasible, as the circuit depth and qubit count remain within the capabilities of current devices.

\subsection{Data Efficiency}
Quantum models are theoretically predicted to possess higher expressibility per parameter than classical neural networks, potentially allowing them to generalize from smaller datasets. Our results support this hypothesis. The classical baseline \cite{manti2025github} utilized a dataset of $>20,000$ waveforms to train the DAE+CNN pipeline. In contrast, our VQC was trained on a balanced subset of approximately $11,000$ events. 
Achieving equivalent AUC (0.98) with nearly half the training data suggests that the VQC ansatz is a more data-efficient learner for this topology of signal, capturing the relevant physical discriminants (rise-time, current collection topology) without requiring the massive over-parameterization typical of classical deep learning.

\subsection{Quantum-Inspired Classical Computing}
Since this work was performed via state-vector simulation on GPUs, it formally falls under the category of ``Quantum-Inspired" Machine Learning. Even without actual quantum acceleration, the result demonstrates that the mathematical structure of quantum circuits—specifically the mapping of data into a high-dimensional Hilbert space via Amplitude Encoding—acts as a superior feature kernel compared to standard translation-invariant convolution for this type of time-series data. This suggests that ``Quantum-Inspired" models can be competitive with respect to standard DL even on classical hardware, offering a middle ground of high efficiency and low complexity.

\subsection{Future Outlook: The Quantum Native Sensor}
While simulation suffices for 10 qubits, scaling this approach to higher timing resolutions (e.g., 4096 samples $\to$ 12 qubits) becomes exponentially expensive on classical hardware. Future execution on NISQ (Noisy Intermediate-Scale Quantum) hardware will be necessary to scale beyond this limit. 
A common concern with NISQ hardware is noise. However, our reliance on a \textit{Progressive Layer-wise} training strategy is a known mitigation technique for connectivity-limited and noisy devices. By solving the optimization problem in small, manageable steps (adding one layer at a time), we prevent the optimizer from getting lost in the ``Barren Plateaus" of the loss landscape.
Ultimately, this work points toward a ``Quantum Native" instrumentation paradigm: if future quantum sensors can transduce signals directly into quantum states (skipping classical digitization), the downstream processing can be handled by small quantum variational circuits, eliminating the Von Neumann bottleneck of data transfer and enabling real-time, low-power quantum signal processing.
\\ \\ 
Furthermore, the high expressibility of the VQC ansatz suggests its applicability extends beyond binary classification. Future studies will investigate the regression of continuous parameters, specifically the reconstruction of the interaction position (fiducial volume definition) and incident gamma directionality, by exploiting the subtle correlations in the charge collection time-series that standard classical filters often discard.

\section{Conclusion}

In this work, we have presented the first application of Quantum Machine Learning to the challenge of Pulse Shape Discrimination in Broad Energy Germanium detectors. By encoding 1024-sample waveforms directly into a 10-qubit Hilbert space via Amplitude Encoding, we constructed a Variational Quantum Circuit that achieves a \textbf{ROC AUC of 0.98} and a global accuracy of \textbf{97.1\%}, effectively matching the performance of current state-of-the-art classical Deep Learning models.

This result highlights the the extreme parametric efficiency of the quantum approach. While standard classical pipelines require tens of thousands of parameters and computationally expensive pre-processing steps (such as Denoising Autoencoders) to handle microphonic noise, our VQC model achieves equivalent fidelity with only \textbf{302 trainable parameters} and operates directly on \textbf{raw waveforms}. This represents a model compression factor of over two orders of magnitude and obviates the need for explicit noise-filtering steps, as the quantum feature map naturally separates signal from background in the high-dimensional state space.

Furthermore, we demonstrated that the challenge of ``Barren Plateaus" in training deep quantum circuits can be effectively overcome using a Progressive Layer-wise training strategy. This allowed us to train a 10-layer ansatz on classical hardware in under 35 minutes, achieving high Signal Acceptance (98.7\%) crucial for rare-event searches like neutrinoless double-beta decay.

This study serves as a proof-of-principle that ``Quantum-Inspired" models are already competitive for high-precision instrumentation tasks. It lays the groundwork for future detector readout systems where quantum sensors might couple directly to quantum processing units, enabling ultra-low latency, low-power, and high-fidelity event classification in the next generation of nuclear and particle physics experiments.

\section*{Acknowledgments}
“BOOST – Boosted Object and Oriented Space Topologies from VBS@HL-LHC” - CUP I57G21000110007, funded by the PNRR ICSC ``Fundamental Research \& Space Economy" (SPOKE 2), code CN00000013, CUP I53C21000340006 - Mission 4, Component 2, Investment 1.4.
The author gratefully acknowledges the \textbf{UniNuvola} Cloud infrastructure of the University of Perugia and INFN for providing the high-performance computing resources used in this study.
\\ \\ 
The author gratefully acknowledges the VIP-2 collaboration for making the dataset~\cite{ZenodoData} publicly available. This dataset is distributed under the Creative Commons Attribution 4.0 International License (CC BY 4.0).
The analysis, methodology, and conclusions presented in this article are those of the author alone and do not represent an official publication, result, or statement of the VIP-2 collaboration.

\bibliographystyle{iopart-num} 
\bibliography{references}

\end{document}